# Thermodynamics of stacking faults and phase stability in cobalt alloys: A combined computational and experimental study

Zheng Zhong[1]†, Ziqi Cui[2]†, Yu Zhuo[1], Tianyu Yu[1], Jianfeng Cai[3], Kaibo Zou[1], Jiacheng Shen[1], Bowen Huang[1], Zhuoming Xie[4], Huiqiu Deng[3], Yang Yu[5], Hao Zhang[5], Wangyu Hu[1*], Tengfei Yang[1*], Jie Hou[1*]

1. *College of Materials Science and Engineering, State Key Laboratory of Cemented Carbide, Hunan University, Changsha 410082, China;*
2. *Department of Mining and Materials Engineering, McGill University, Montreal H3A 0C5, Canada*
3. *School of Physics and Electronics, Hunan University, Changsha 410082, China*
4. *Key Laboratory of Materials Physics, Institute of Solid State Physics, Chinese Academy of Sciences, Hefei 230031, China*
5. *State Key Laboratory of Cemented Carbide, Zhuzhou, Hunan 412000, China*

†. *These authors contributed equally to this work*

* *Corresponding authors: wyuhu@hnu.edu.cn (Wangyu Hu), yangtengfei@hnu.edu.cn (Tengfei Yang), jiehou@hnu.edu.cn (Jie Hou)*

**Abstract:**

Stacking fault energy dictates phase stability and deformation behavior in Co alloys and WC-Co cemented carbides, yet a quantitative assessment of alloying effects at finite temperatures remains poorly established. By integrating first-principles thermodynamics with microstructural characterization, we provide a rigorous evaluation of these influences across atomic and macroscopic scales. We show that stacking fault energetics at 0K for transition metal solutes are primarily governed by atomic misfit volume. While 4d and 5d elements follow a consistent linear trend, specific 3d solutes exhibit significant deviations due to non-negligible magnetic contributions. By incorporating phonon, electronic, longitudinal spin-fluctuation, and magnetic free-energy contributions, the model accurately captures the fcc-hcp transformation and quantifies how diverse solutes modulate the phase landscape. We demonstrate that V, Ni, Fe, Mo, and W lower the transformation temperature by stabilizing fcc phase, while Cr and C exhibit the opposite effect, consistent with experimental phase diagrams. Furthermore, microscopic analysis confirms that higher W content dissolved in the Co suppresses stacking-fault formation by elevating the stacking fault energy at finite temperatures. This work clarifies the physical mechanisms by which alloying regulates stacking fault energy and phase stability in Co-based systems, providing guidance for the design of Co-based alloys and WC-Co cemented carbides.

## 1. Introduction

Cobalt (Co) and Co-based alloys exhibit excellent high-temperature mechanical properties, outstanding corrosion resistance, and superior wear resistance, making them indispensable in critical applications such as turbine blades, biomedical implants, and cemented carbides [1-4]. Across a wide spectrum of applications, ranging from classical WC-Co composites and Cr alloyed medical alloys to Co-Ni-based superalloys, various transition metals (TM)—such as Cr, Ni, W, Ta, Ti, and Re—have been extensively utilized to enhance performance through solid solution and precipitation strengthening mechanisms [5, 6]. Furthermore, within the specific domain of WC-Co cemented carbides, TM additions serve a highly specialized role. Solutes such as V, Cr, Nb, Ta, and Ti are routinely added as grain-growth inhibitors to suppress WC grain coarsening, which critically improves microstructural stability and service performance [7, 8].

Realizing the potential of multi-component Co-based systems requires a fundamental understanding of how alloying elements influence key microscopic structural parameters of the Co matrix. Among these, the stacking fault energy (SFE) is particularly critical, linking defect energetics at the atomic scale to macroscopic mechanical behavior [9]. SFE strongly modulates deformation modes such as dislocation slip and mechanical twinning, as well as the phase transformation. Low SFEs are typically associated with transformation-induced plasticity (TRIP) and twinning-induced plasticity (TWIP), while high SFEs are generally related to dislocation-mediated plasticity [10, 11]. In high-Mn steels, for instance, Al and Si addition has been shown to increase and decrease SFE, respectively, thereby enabling control over TRIP and TWIP behaviors [12]. In pure Co, the martensitic transformation between the hexagonal close-packed (hcp) $\varepsilon$ phase and the face-centered cubic (fcc) $\gamma$ phase occurs near 700 K and is closely related to the stacking fault energy [13]. Since the ($\gamma \rightarrow \varepsilon$) transformation proceeds through changes in the close-packed stacking sequence, alloying-induced variations in SFE can strongly affect the relative stability of the two phases and the transformation temperature. For instance, doping Co with 6 at.% Fe or 30 at.% Ni is known to reduce the transformation temperature to room temperature [14, 15]. In cemented carbide systems, Eizadjou et al. [16] showed that Ru addition decreases the stacking fault energy of the Co binder phase, thereby promoting the martensitic transformation. More recently, Zhou et al. [17] further demonstrated that increasing the solubility of W and C in the Co phase can raise the stacking fault energy, suppress abnormal WC grain growth, and increase the retained fcc Co fraction at room temperature.

These insights underscore the central role of SFE as a design parameter for tailoring phase stability and deformation behavior in Co-based alloys. However, accurate characterization of the SFE and $\gamma/\varepsilon$ phase stability in practical Co-based materials remains challenging. On the one hand, processing routes such as rapid solidification, quenching, additive manufacturing, and other non-equilibrium heat treatments frequently produce supersaturated solid solutions. Conventional phase diagrams, however, primarily describe equilibrium phase stability and therefore provide limited thermodynamic information on the $\gamma/\varepsilon$ stability and SFE of such metastable or non-equilibrium solid solutions. On the other hand, despite advances in experimental characterization techniques, precise determination of SFE in metastable metallic systems remains difficult. This difficulty partly arises from the intrinsic limitations of indirect measurement methods, which often constrain the SFE to positive values because of their underlying model assumptions [18, 19]. By contrast, atomistic simulations can easily circumvent these constraints, positioning computational approaches as a vital complementary route for SFE quantification. Extensive density functional theory (DFT) investigations have been conducted on SFE calculations for Co and Co-based alloys, yet marked inconsistencies persist across reported values. Notably, Achmad et al. [20-22] reported a 0 K SFE of 0.15 mJ/m$^2$ for pure Co with a positive temperature dependence—a finding that contradicts the experimentally observed stability of the hcp phase at ambient temperatures. Tian et al. [23] attributed this discrepancy to the introduction of an additional stacking fault in Achmad et al.'s model, and after correction, obtained a negative SFE of -106.2 mJ/m$^2$, more consistent with thermodynamic expectations. In their subsequent SFE calculations for Co-based alloys, they employed the coherent potential approximation (CPA). While this approach offers substantial computational efficiency for modeling random alloys, its tendency to average out local atomic environments raises questions about its accuracy relative to explicit supercell methods.

In addition to the SFE, similar inconsistencies appear in DFT-predicted transformation temperatures for pure Co. Wang et al. [24] estimated a transition at 650 K accounting for phonon and electronic excitations, whereas Lizárraga et al. [25] reported 1100 K under similar conditions. To address this discrepancy, the latter study further incorporated magnetic interactions and longitudinal spin fluctuations, eventually yields a transformation temperature of 825 K. Nonetheless, systematic deviations from experimental values (~700 K) remain unresolved. Crucially, thermodynamic analyses of Co-based alloys predominantly rely on CALPHAD methods for phase transition predictions. However, the pure-element databases used in CALPHAD typically employ polynomial models fitted to experimental thermodynamic data, which lack physical interpretability and may become unreliable when extrapolated beyond the fitted range, leading to inconsistencies between low- and high-temperature data [26].

Despite these extensive efforts, the current understanding of the SFE in Co-based alloys remains fragmented, with particular scarcity in quantitative assessments of temperature-dependent SFE behavior. To address this issue, the present study systematically investigates the SFE and its temperature dependence in Co-based systems, with the aim of providing guidance for the design and optimization of Co-based alloys and cemented carbides. In this work, first-principles thermodynamic calculations are employed to systematically investigate the effects of alloying elements on the intrinsic stacking fault energy and phase stability of Co-based alloys. The results show that, at 0 K, the stacking fault energy of Co-based alloys is governed by the combined effects of solute misfit volume and magnetic contributions. By further incorporating phonon, electronic, and magnetic free-energy contributions, a finite-temperature thermodynamic description of the stacking fault energy and phase stability of pure Co and representative dilute Co-based alloys is established. TEM/EDS characterizations of WC-Co cemented carbides are conducted, which show that W solutes suppress stacking-fault formation in Co, confirming our computational predictions. These findings provide a systematic understanding of the physical mechanisms by which alloying regulates stacking fault energy and phase stability in Co-based systems, and offer theoretical guidance for microstructural control and deformation-behavior design in Co-based alloys and WC-Co cemented carbides.

## 2. Methods

*2.1 Explicit model for intrinsic stacking fault energy calculation*

In fcc structures, dislocation slip typically occurs along the $< 110 >$ directions on $\{111\}$ planes. Under favorable conditions, a perfect dislocation of type $< 110 >/2$ can lower its energy by dissociating into two $< 112 >/6$ Shockley partial dislocations (Fig. 1b), forming a stacking fault ribbon between them. Within the framework of DFT, such stacking fault structures can be constructed by applying a shear displacement to the supercell. Fig. 1c illustrates the ideal fcc stacking sequence ABCABC. An intrinsic stacking fault (ISF) is generated by applying a displacement vector of $\frac{1}{6}[11\bar{2}]$ to the simulation cell, while keeping the atomic Cartesian coordinates fixed. As a result, the stacking sequence in the shaded region of Fig. 1d then transforms into ABAB, locally adopting a hcp structure and forming an ISF. Based on this model, the free energy of the ISF under given temperature conditions can be computed following:

$$\gamma_{ISF} = \frac{E_{ISF} - E_{fcc}}{A}, \quad (1)$$

where $E_{ISF}$ and $E_{fcc}$ denote the energy of the ISF and perfect configurations, respectively. $A$ is the stacking fault area included in the simulation box. This explicit treatment captures local relaxation around the fault plane and is therefore well suited to resolving defect-induced changes in atomic environment, magnetic moments and solute–fault interactions. Its main limitation, however, is the substantial computational cost associated with large supercells and reduced symmetry, particularly when finite-temperature free-energy contributions are included.

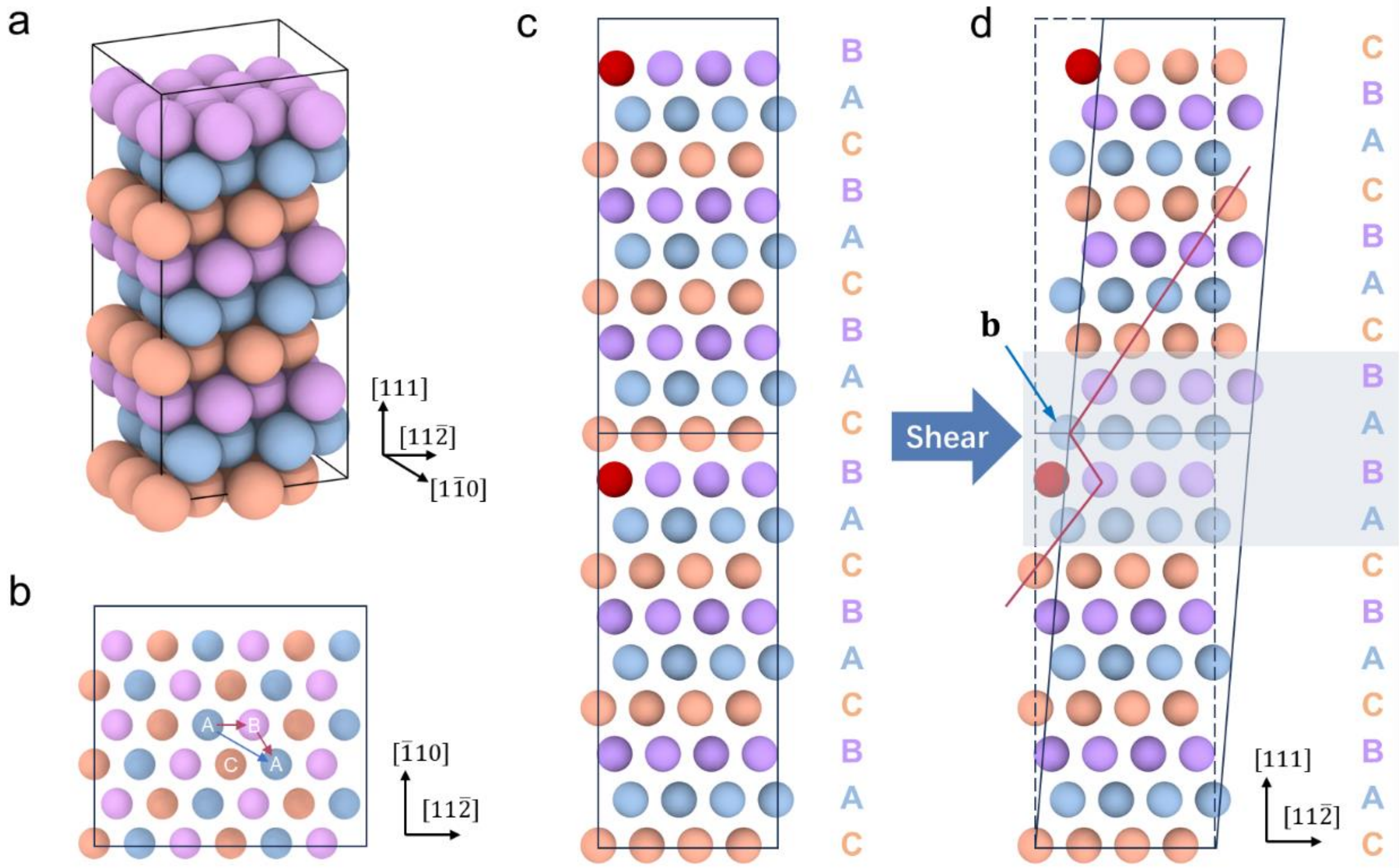

**Fig. 1.** Atomistic model of intrinsic stacking fault in fcc structures. (a) Ideal fcc lattice illustrating ABC stacking sequence, with atoms in A (blue), B (purple), and C (tan) layers. (b) Schematic of the $(111)$ plane, with red and blue arrows indicating Burgers' vectors for Shockley partial dislocations and perfect dislocations, respectively. (c) $[\bar{1}10]$ crystallographic plane of an ideal fcc structure. The upper half is a periodic image of the lower half. The red atom marks the TM doping site. (d) ISF configuration generated by applying a displacement vector of $\frac{1}{6}[11\bar{2}]$ to the simulation cell, while keeping the atomic Cartesian coordinates fixed. Blue region highlights faulted AB stacking sequence exhibiting local hcp phase atomic arrangement.

### *2.2 Free-energy framework for temperature-dependent ISFE calculation*

To access the temperature dependence of the intrinsic stacking-fault energy more efficiently, we additionally employ the axial next-nearest-neighbour Ising (ANNNI) approximation, in which the ISF is treated as a local hcp-like stacking perturbation on the fcc (111) planes [27, 28]. Within this first-order description, the ISF free energy is directly related to the bulk free-energy difference between the fcc and hcp phases:

$$\gamma_{ISF} = \frac{F_{hcp}(V,T) - F_{fcc}(V,T)}{A'}, \tag{2}$$

where $A' = \sqrt{3}a_0^2/4$ is the normalized stacking fault area, with $a_0$ being the lattice constant of the fcc structure, and $F_i(V,T)$ (where i = fcc, or hcp) denote the corresponding per-atom Helmholtz free energy. The Helmholtz free energy at a given volume $V$ and temperature $T$ is decomposed into multiple contributions as follows:

$$F_i(V,T) = E_{st}(V) + F_{vib}(V,T) + F_{el}(V,T) + F_{mag}(V,T) + F_{lsf}(V,T), \tag{3}$$

where $E_{st}(V)$ denotes the static energy of the supercell, while $F_{vib}(V,T)$, $F_{el}(V,T)$, $F_{mag}(V,T)$, and $F_{lsf}(V,T)$ represent the contributions to the free energy from phonons, electronic excitations, magnetism, and longitudinal spin fluctuations, respectively. Within the quasi-harmonic approximation (QHA), the vibrational free energy $F_{vib}$ can be obtained from the phonon spectrum and is expressed as [29]:

$$F_{vib} = \frac{1}{2}\sum_{qv} \hbar\omega_j(\boldsymbol{q}, v) + k_B T \sum_{qv} \ln\left\{1 - \exp\left[-\frac{\hbar\omega_j(\boldsymbol{q}, v)}{k_B T}\right]\right\}, \tag{4}$$

where $k_B$ is the Boltzmann constant, and $\hbar\omega_j(\boldsymbol{q}, v)$ is the phonon energy corresponding to wave vector $\boldsymbol{q}$ and mode $v$. The first term corresponds to the zero-point energy (ZPE).

The electronic contribution, $F_{el}$ in Eq. (3), is obtained using finite-temperature DFT combining Fermi-Dirac distribution. While this term is negligible at low temperatures, it becomes increasingly significant at elevated temperatures and must be included for accurate thermodynamic modelling. In Co-based systems, magnetic contributions to the free energy are also non-negligible. Many existing studies of temperature-dependent ISF energies employ empirical thermodynamic models to approximate these magnetic effects. However, such approaches may introduce substantial errors when capturing the subtle thermodynamic behavior of alloyed Co systems [21, 30]. To ensure accurate description on free energies, we adopted classical Monte Carlo (cMC) simulations based on the Heisenberg model [31-33]. Neglecting uniaxial anisotropy and external field contributions, the Hamiltonian simplifies to:

$$\mathcal{H} = -\sum_{i<j} J_{ij} \mathbf{S}_i \cdot \mathbf{S}_j, \tag{5}$$

where $J_{ij}$ is the exchange interaction constant (positive for ferromagnetic coupling, negative for antiferromagnetic) between atom $i$ and $j$, and $\mathbf{S}$ is the spin quantum number. The local magnetic moment $M_0$ is related to $\mathbf{S}$ by $M_0 = g\mu_B\mathbf{S}$, where $g$ is the Landé g-factor and $\mu_B$ is the Bohr magneton. From Eq. (5), the magnetic heat capacity $C_V$ can be obtained via cMC simulations, and the magnetic entropy is then calculated as:

$$S_{mag} = \int_0^T \frac{C_V(T)}{T} dt. \tag{6}$$

Eqs. (5-6) capture the contribution of transverse spin fluctuation to the free energy. To incorporate longitudinal spin fluctuations (lsf), we adopt the approach proposed by Lizárraga et al. [25], performing fixed spin moment calculations to obtain the spin density distribution (SDD) as a function of temperature. The SDD follows the Boltzmann distribution:

$$p_i = \frac{1}{Z} \exp\left[-\frac{E_i}{k_B T}\right], \tag{7}$$

where $E_i$ is the energy of the system under a given local magnetic moment $M_0$, and $Z$ is the partition function. The entropy contribution from lsf can be computed as:

$$S_{lsf} = -k_B \int \mathcal{P}(M_0)\, ln(\mathcal{P}(M_0))\, dM_0, \tag{8}$$

where $\mathcal{P}(M_0)$ is a continuous probability distribution function obtained by interpolating the discrete values of $p_i$. With the entropy values from Eqs. (6) and (8), the free energy contributions $F_{mag}$ and $F_{lsf}$ can be evaluated using $F = U - TS$. By incorporating all the above contributions, we are able to determine the free energy of the system and thereby accurately evaluate the temperature dependence of the stacking fault energy.

*2.3 General computational parameters*

We performed DFT calculations using the Vienna Ab initio Simulation Package (VASP) [34, 35]. The projector augmented wave (PAW) method was employed to describe electron-ion interactions, with the Perdew-Burke-Ernzerhof (PBE) exchange-correlation functional under the generalized gradient approximation (GGA) framework [36, 37]. Spin polarization was included in all calculations to accurately characterize magnetic properties. In the static calculations, the plane-wave cutoff energy was set to 350 eV, with convergence criteria of $10^{-7}$ eV for the total energy and 0.01 eV/Å for atomic forces. For the thermodynamic calculations, the cutoff energy was increased to 400 eV, with stricter convergence criteria of $10^{-8}$ eV for energy and $10^{-6}$ eV/Å for forces. Partial occupancies were handled using the first-order Methfessel-Paxton scheme [38] with a smearing width of 0.1 eV. Brillouin zone sampling was performed using the Gamma-centered Monkhorst-Pack scheme. The explicit stacking fault energy was calculated using the model described in Eq. (1), where the fcc supercell shown in Fig. 1a-1b was adopted with a 4×2×2 k-point mesh. For the temperature-dependent stacking fault energy analysis, the ANNNI model [Eq. (2)] was applied to 3×3×3 orthorhombic supercells of both fcc and hcp phases, each containing 108 atoms. For hcp, the orthorhombic supercell was defined with lattice vectors along $[11\bar{0}0]$, $[1\bar{1}\bar{2}0]$, and $[0001]$. Monkhorst-Pack k-point meshes of 3×3×3 and 5×3×3 were used for the fcc and hcp structures, respectively. Phonon and electronic contributions were computed within the density functional perturbation theory (DFPT) framework using the Phonopy software package [39]. In the magnetic calculations, the exchange coupling constants $J_{ij}$ were truncated at a cutoff distance of 10 Å, and Monte Carlo simulations under the Heisenberg model were performed using the Vampire software package [40].

### *2.4 Experimental details*

Microstructural characterization and compositional analysis were performed on two commercial YK05 WC-Co cemented carbide samples containing 6 wt.% Co, supplied by Zhuzhou Cemented Carbide Group Co., Ltd. Transmission electron microscopy (TEM) observations and energy-dispersive X-ray spectroscopy (EDS) analyses were carried out using an aberration-corrected Thermo Fisher Scientific Themis Z microscope operated at 300 kV. The two investigated conditions were: (i) an as-sintered sample sintered at 1450 °C; and (ii) a quench-tempered sample, which was oil-quenched from approximately 1150 °C after sintering and subsequently tempered at 400 °C for 2 h.

## 3. Results

### *3.1. ISFE of Co-based Alloys at 0 K*

We begin by examining the fundamental properties of pure Co at 0 K. As summarized in Table 1, our calculated lattice constants and atomic volume show excellent agreement with both previous DFT results and experimental values. The only notable deviation lies in the intrinsic stacking fault energy (ISFE). Our calculated ISFE for fcc Co is -104.53 mJ/m$^2$, which deviates significantly from the 0.15 mJ/m$^2$ obtained by Achmad et al. [20], but closely matches the value of -106.2 mJ/m$^2$ reported by Tian et al. [23] This agreement lends further support to Tian et al.'s assertion that the discrepancy in Achmad et al.'s result stems from the inadvertent introduction of an additional stacking fault in their supercell model.

**Table 1. Comparison of calculated basic physical properties of Co with theoretical and experimental data.**

| Property | This work | Cal. | Exp. |
|---|---|---|---|
| Lattice constant (hcp, a, Å) | 2.492 | 2.502 [24]<br>2.518 [41] | 2.505 [42]<br>2.507 [43] |
| Lattice constant (hcp, c/a) | 1.616 | 1.600 [24]<br>1.618 [41] | 1.631 [42]<br>1.602 [43] |
| Lattice constant (fcc, a, Å) | 2.483 | 2.498 [24];<br>2.512 [41] | 2.506 [42]<br>2.503 [44] |
| Atomic volume $V_0$ (hcp, Å$^3$) | 10.81 | 10.83 [24]<br>11.13 [41] | 11.06 [42]<br>10.02 [45] |
| Atomic volume $V_0$ (fcc, Å$^3$) | 10.84 | 11.02 [24]<br>11.16 [41] | 11.08 [42]<br>11.09 [44] |
| ISFE (fcc, mJ/m$^2$) | -104.53 | -106.2 [23]<br>0.15 [20] | |

Following the analysis of pure Co, we next examine the behavior of TM solutes in the Co matrix. The substitutional dissolution energy was calculated as

$$E_{\text{sol}} = E(\text{Co}_{N-1}X) - E(\text{Co}_N) + \mu_{\text{Co}} - \mu_X, \tag{9}$$

where $E(\text{Co}_{N-1}X)$ and $E(\text{Co}_N)$ are the total energies of the Co supercell with and without one substitutional TM solute $X$, respectively, and $\mu_{\text{Co}}$ and $\mu_X$ denote the chemical potentials of Co and the solute element in their corresponding stable elemental phases. Fig. 2 illustrates the distribution of dissolution energy for 27 TM elements. Most dopants exhibit positive solution energies, indicating their poor thermodynamic stability in Co matrix and limited solid solubility. Notably, Ti exhibits the lowest solution energy among all dopants, suggesting the most favorable solid solution effect, while Hg shows the highest dissolution energy. Notably, 4d and 5d TM elements generally demonstrate higher dissolution energies compared to 3d counterparts. This trend is primarily attributed to their larger atomic radii, which induce significant lattice mismatch and consequently higher elastic strain energy.

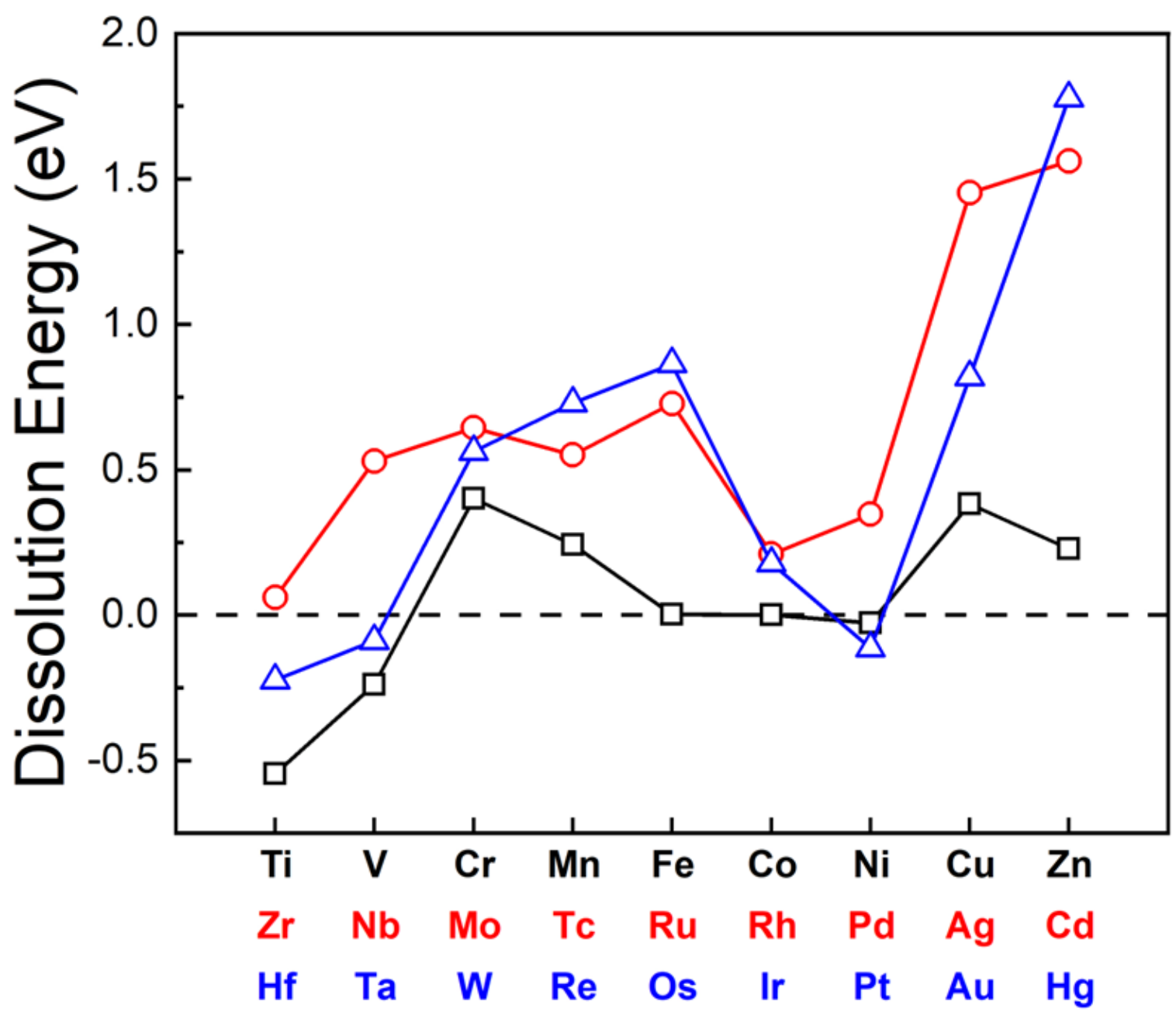

**Fig. 2.** Dissolution energies of TM solutes in fcc Co.

Building on the assessment of solution behavior, we further investigate the influence of dopants on SFE, a key microscopic mechanism governing phase stability in the Co matrix. Fig. 3a shows the calculated ISFE for Co-based alloy systems at 0 K, with one TM atom replacing a Co atom within the stacking fault plane (Fig. 1c). Dopants that lower the ISFE below the dashed reference line are predicted to stabilize hcp phase, whereas those that increase the ISFE favor fcc phase stabilization. Among the dopants, Cr and Re exhibited the most pronounced effect in lowering ISFE, suggesting their strong potential for enhancing hcp phase stability at low temperatures.

For 4d/5d elements, the ISFE variation follows a distinct parabolic trend with respect to atomic number, in contrast to the irregular fluctuations observed across the 3d series elements. Notably, Mn and Fe deviate from the general group trend, showing anomalous ISFE values. To elucidate the origin of this anomaly, we analyzed the atomic misfit volumes for these TM solutes in fcc Co. As shown in Fig. 3b, the misfit volume follows a similar parabolic trend in general, with pronounced deviations for Mn and Fe, mirroring the behavior seen in the ISFE. This correlation highlights the strong influence of volumetric strain on stacking fault energetics and underscores the critical role of lattice distortion in governing phase stability in Co-based alloys.

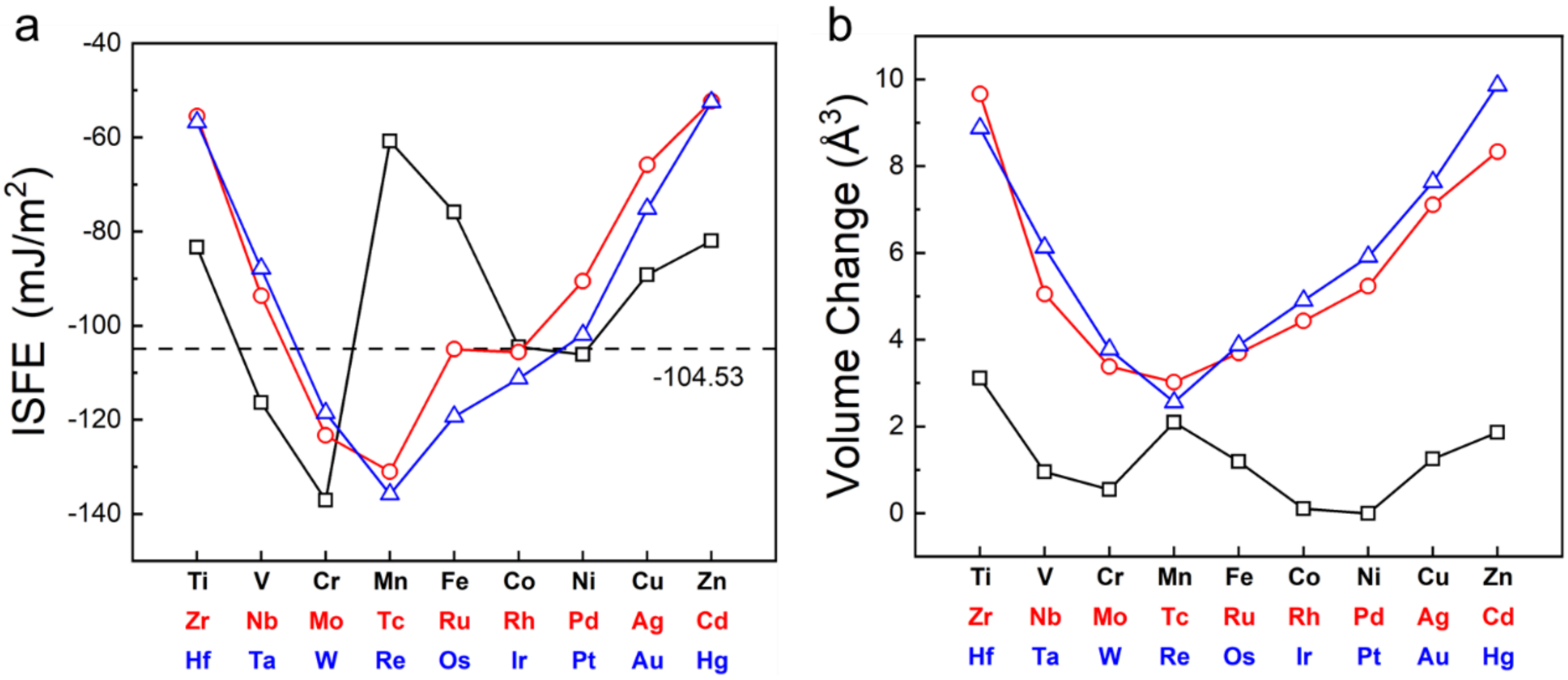

**Fig. 3.** (a) ISFE of Co-based binary alloys at 0 K. The dashed line indicates the ISFE of pure Co. (b) Relative volume changes upon substitutional doping of TM solutes in fcc Co.

To quantify this correlation, we conducted a linear regression analysis between misfit volume and stacking fault energy, as shown in Fig. 4a. The results show that the ISFE generally increases with increasing misfit volume, which means large TM atoms inhibit ISF formation. Since stacking faults in fcc crystals resemble local hcp structures, this trend suggests that oversized solutes preferentially stabilize the fcc phase of Co over the hcp phase. This can be rationalized by the relatively more open atomic packing of the fcc structure (see Table 1), which better accommodates large solute atoms with minimal lattice distortion. Similar observations have been reported in Nb-based alloys, where Shi et al. [46] suggested that atomic radius and valence electron concentration serve as key indicators for evaluating the ISFE, further confirming the critical role of misfit volume in influencing the ISFE.

A closer examination of Fig. 4a reveals that the ISFE exhibits a strong linear correlation ($R^2$ close to 1) with misfit volume for 4d and 5d TM solutes, indicating that their ISFEs is predominantly governed by geometric factors. In contrast, the 3d series shows significant deviations from this trend, suggesting a more complex interplay of underlying mechanisms. In light of prior studies that associate the ISFE with electronegativity and bonding character, we extended our analysis to these electronic descriptors to evaluate their relevance in explaining the anomalous behavior within the 3d series. Specifically, we examined Mn, Tc, and Re, which exhibit nearly identical misfit volumes and electronegativities, yet display markedly different ISFE. To further investigate this discrepancy, we analyzed the Electron Localization Function (ELF) and Density of States (DOS) for these systems. The bonding characteristics of all three systems were found to exhibit typical metallic bonding without significant differences (see Supplementary Fig. S1), indicating that neither electronegativities nor bonding character alone can fully explain the ISFE evolution in the 3d series.

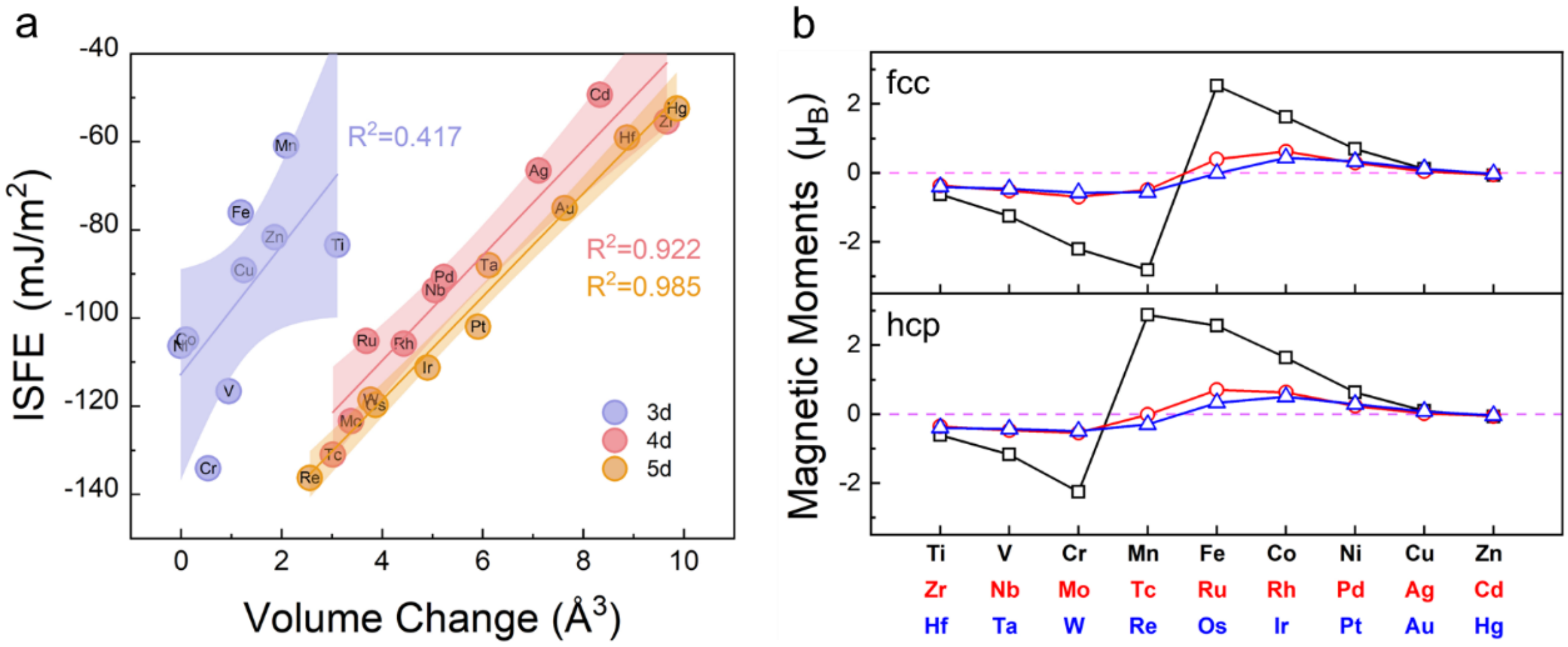

**Fig. 4.** (a) Correlation between misfit volume and ISFE for Co-based systems. Separate linear regressions were performed for 3d, 4d, and 5d elements. The shaded area represents the 95% confidence interval. (b) Local magnetic moments of TM dopants in Co. The upper panel corresponds to the fcc phase, and the lower panel to the hcp phase.

Upon further examination, we find certain TM solutes (most notably Mn, Fe, and Cr) demonstrate particularly prominent deviation from the linear relationship. Given the well-known magnetic nature of these elements, we further analyzed their local magnetic moments in the Co matrix. As shown in Fig. 4b, we find 4d and 5d TM solutes exhibit comparably negligible magnetic moments, indicating limited magnetic contributions to their ISFEs and reinforcing the dominance of geometric effects in these systems. By contrast, the 3d series displays a more nuanced trend. Low-magnetic dopants (e.g., Cu, Zn, Ni) closely follow the misfit-volume scaling relationship, whereas magnetic elements deviate significantly. Specifically, elements with negative (in comparison to Co matrix) magnetic moments tend to fall below the fitted baseline, while those with positive moments lie above. Mn stands out as particularly anomalous, exhibiting a sign reversal of magnetic moment between fcc (+3 μB) and hcp (-3 μB) Co. This leads to an ISFE shift far exceeding that predicted by volumetric considerations alone.

This anomalous behavior originates from the near-half-filled 3d electronic configuration of Mn, which renders its magnetic state intrinsically sensitive to the local crystal structure [47]. Similar structure-sensitive magnetic behavior of Mn has also been reported in bcc Fe [48]. The distinct coordination environments in fcc and hcp structures induce spin-state switching and magnetic coupling inversion. According to the ANNNI model, ISFE approximates the free energy difference between hcp and fcc phases, normalized by the stacking fault area. Thus, magnetically driven amplification of free energy differences emerges as a critical mechanism modulating the ISFE behavior of elements like Mn, explaining their anomalous characteristics among the 3d transition metals.

Note that the above calculations were performed with TM solutes substituting a Co atom in the two atomic layers adjacent to the intrinsic stacking fault (as illustrated in Fig. 1d). However, in realistic alloy systems, such proximity is not always guaranteed, and TM solutes are more likely to be randomly distributed with respect to the stacking fault plane. To obtain a more accurate estimate of the ISFE, it is therefore essential to evaluate the influence of solutes placed at various atomic layers and average their impact accordingly.

To this end, we evaluated the position-dependent ISFE for representative TM solutes, including W and Cr (commonly used in cemented carbides) and Cd (due to its most pronounced ISFE-increasing effect). As illustrated in Fig. 5a, the impact of the solute is highly sensitive to its proximity to the fault plane. The effect is maximized when the solute is positioned directly within the stacking fault region (i.e., the 5th and 6th layers). However, this influence diminishes rapidly as the solute moves away, becoming negligible beyond two atomic layers where the ISFE essentially fluctuates around the bulk value of pure Co.

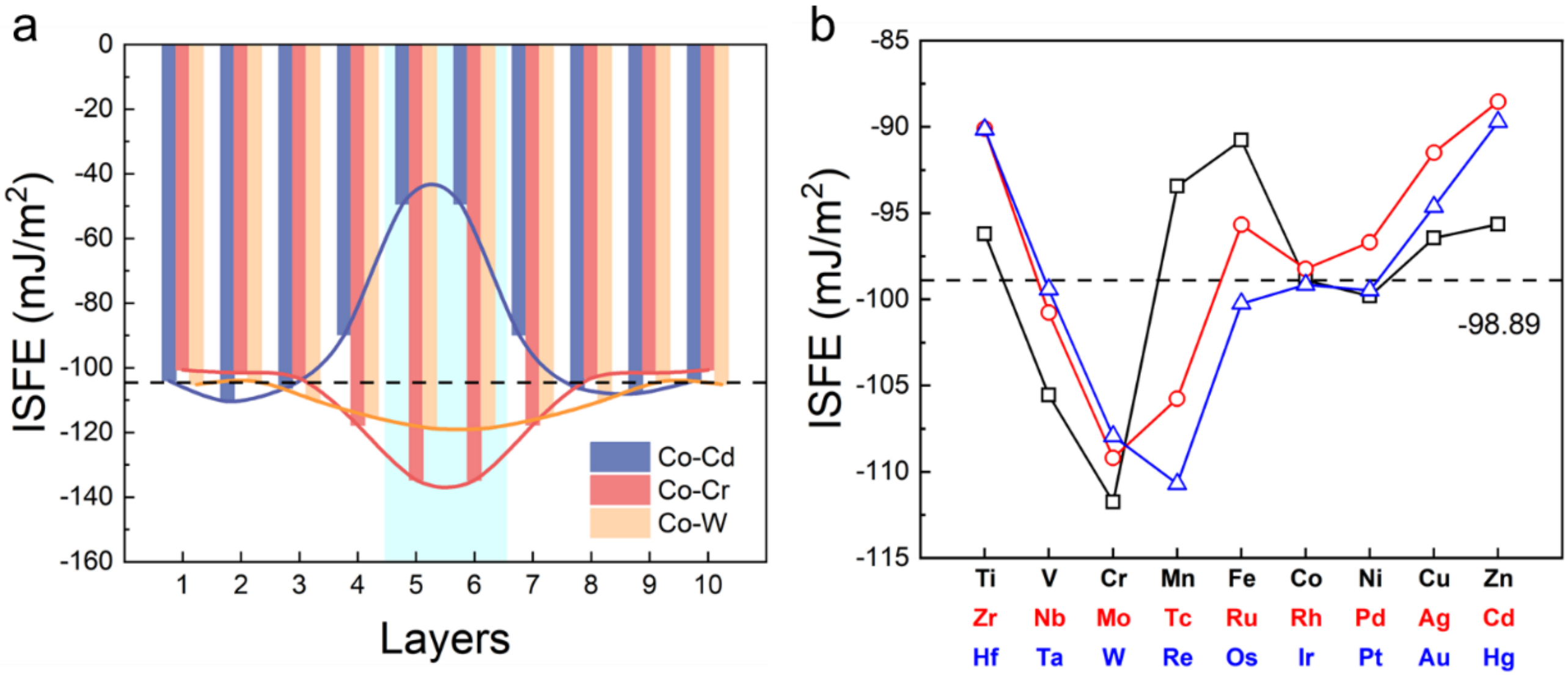

**Fig. 5.** (a) Variation in ISFE with doping position for W, Cr, and Cd in the explicit stacking fault model. The 5th and 6th layers correspond to the stacking fault plane (indicated by the blue region), and the 10th layer represents the periodic replicate of the 1st layer. The black dashed line indicates the ISFE of pure Co. (b) ISFE values of Co-based alloys predicted using the first-order approximation of the ANNNI model.

Before transitioning to finite-temperature thermodynamics, we validated the ANNNI model's baseline accuracy at 0 K (Fig. 5b). The ANNNI predictions successfully capture the compositional trends of the explicit model, yielding an ISFE of -98.89 mJ/m$^2$ for pure Co, consistent with the value from the explicit model (-104.53 mJ/m$^2$). For alloyed systems, numerical deviations arise from differing spatial treatments: the ANNNI framework inherently assumes a homogeneous solute distribution, whereas our explicit baseline localized solutes precisely at the fault plane. This discrepancy can be effectively reconciled by averaging the position-dependent explicit ISFEs (Fig. 5a), which brings the ANNNI predictions within a 5 mJ/m$^2$ margin of the explicit results. Since this static deviation is minor relative to the substantial ISFE shifts induced by thermal effects, the ANNNI model provides a robust framework for the subsequent finite-temperature analysis.

**Table 2. Comparison of ISFE (mJ/m$^2$) calculated using the explicit model and the ANNNI model for Co-based alloys.**

| Model | Pure Co | Co-W | Co-Cr | Co-Cd |
|---|---|---|---|---|
| Explicit | -104.53 | -111.17 | -112.78 | -90.17 |
| ANNNI | -98.89 | -107.94 | -111.75 | -88.55 |

*3.2. ISFE of Co-based Alloys at finite temperatures*

The above results highlight the intrinsic link between ISFEs and the fcc-hcp phase transition in cobalt. To bridge this zero-temperature baseline with finite-temperature thermodynamic stability, we first examine the thermodynamic stability of pure Co. Following Eq. (3), the total Helmholtz free energy was explicitly decomposed into discrete contributions: phonon vibrations ($F_{vib}$), electronic excitations ($F_{el}$), longitudinal spin fluctuations ($F_{lsf}$), and magnetic effects ($F_{mag}$).

Figure 6a illustrates how the transition temperature from the fcc to the hcp phase of pure Co shifts as successive free energy components are incorporated. Phonon vibrations act as the primary driving force for stabilizing the high temperature fcc phase, and this description is validated by excellent agreement with experimental phonon spectra (Supplementary Fig. S2). However, relying solely on vibrational entropy substantially overestimates the transition temperature, which indicates that achieving quantitative accuracy requires additional fundamental excitations. Incorporating electronic excitations becomes prominent above 400 K and lowers the transition temperature to 930 K (Supplementary Fig. S3). By contrast, longitudinal spin fluctuations provide only a marginal correction of roughly 90 K. Finally, the magnetic contribution exerts a substantial effect and decreases the transition temperature by a further 180 K. Since the estimated Curie temperature of 1250 K (Supplementary Fig. S4b) lies well above the fcc–hcp transition temperature, the system remains ferromagnetically ordered, allowing paramagnetic effects to be safely neglected. Integrating all thermodynamic terms yields a predicted transition temperature of 660 K, in close agreement with the experimental value of 690 K. Ultimately, this rigorous validation confirms that the integrated free energy framework can accurately describe the temperature dependent phase transition behavior of Co.

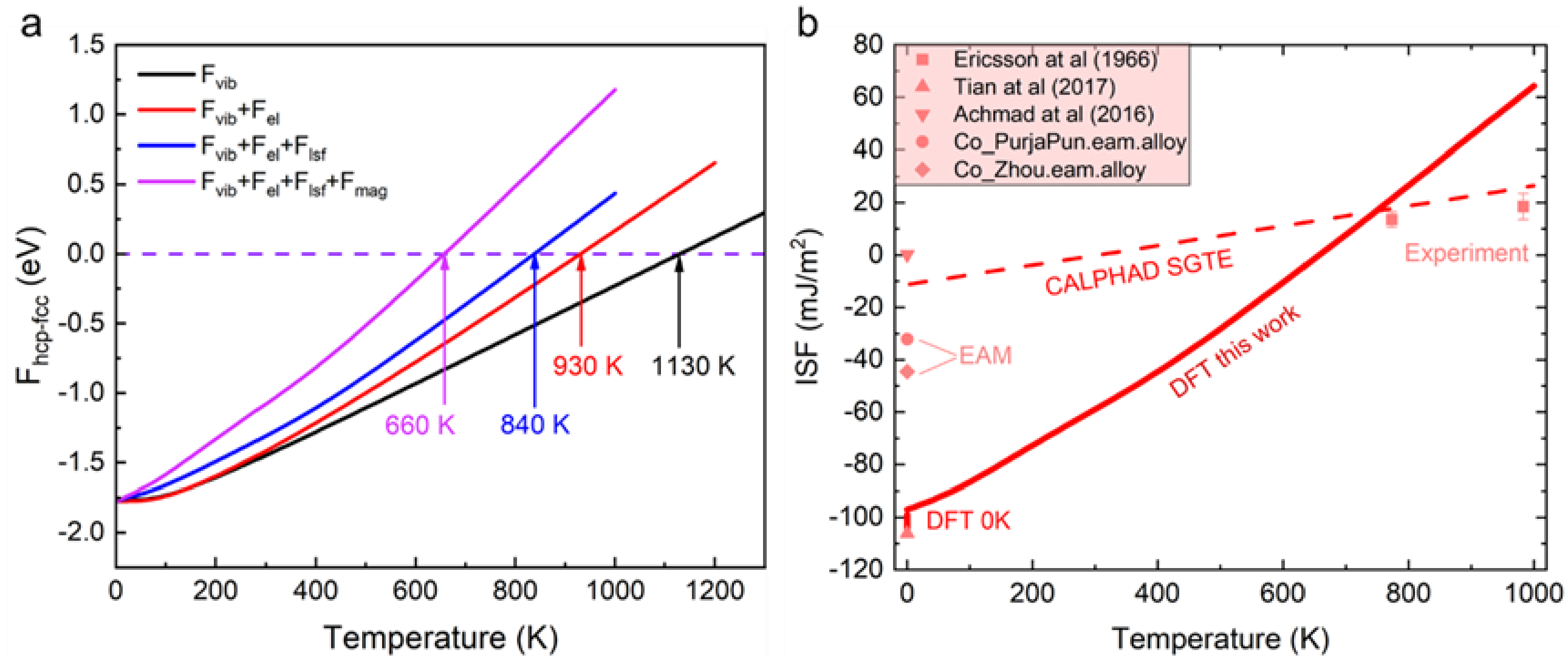

**Fig. 6.** (a) Temperature evolution of the free energy difference ($F_{hcp-fcc}$) for pure cobalt. The dashed line marks the phase transition temperature at $\Delta F = 0$. Black, red, blue, and purple curves represent contributions from vibrational ($F_{vib}$), electronic excitations ($F_{el}$), longitudinal spin fluctuation ($F_{lsf}$), and magnetic ($F_{mag}$) free energies, respectively. Incorporating all thermodynamic contributions, the calculated phase transition temperature of pure Co is 660 K. (b) Temperature-dependent ISFE for pure cobalt.

Next, we further evaluate the reliability of our theoretical framework by benchmarking it against available experimental and empirical data. Because direct measurements in pure Co are scarce and often limited by indirect methodologies, Fig. 6b compiles our DFT-derived temperature-dependent ISFE predictions (obtained via the ANNNI model) alongside CALPHAD-based assessments [21] and the limited experimental values. All approaches reveal a consistent trend of increasing ISFE with temperature. Notably, our DFT results closely match experimental data at intermediate temperatures (e.g., 773 K), but exhibit a steeper temperature dependence. This deviation may reflect the idealized conditions of DFT, while experimental measurements can be influenced by factors such as dislocation density and residual stress fields, which suppress the apparent ISFE. Furthermore, we evaluated the 0 K ISFE of pure Co using two widely used embedded atom method (EAM) potentials developed by Pun [49] and Zhou [50]. The resulting values, -32.12 mJ/m$^2$ and -44.52 mJ/m$^2$ respectively, showing substantial deviation from DFT values (-104.53 mJ/m$^2$), suggesting that these empirical potentials lack sufficient fidelity to accurately capture stacking fault energetics. Together, these comparisons underscore the importance of first-principles accuracy when modeling defect thermodynamics in Co-based systems.

Building upon our pure Co validation, we extended the thermodynamic framework to Co-based alloys. Note since longitudinal spin fluctuations demonstrated a minor thermodynamic contribution (Fig. 6a) alongside substantial computational costs, we used the pure Co values for this specific term across all alloy simulations. Following this strategy, we next examined the temperature dependence of ISFE across a representative subset of common alloying elements, specifically Cr, Mn, Fe, Ni, W, Mo, and the interstitial solute C. It is worth noting that C, as an interstitial solute, introduces high configurational freedom, significantly increasing the computational workload. To mitigate this, calculations for C-alloyed Co were performed using a reduced 54-atom supercell at a concentration of 1.85 at.%, with ISFE values linearly scaled to 0.93 at.% to ensure a consistent comparative analysis with other substitutional alloys.

Fig. 7 shows the thermal evolution of the ISFE across the investigated dilute Co alloys, with the blue shaded region representing the cumulative contributions from vibrational, electronic, and magnetic free energies. A universal monotonic increase in ISFE with temperature is observed across all systems, indicating that thermal activation consistently drives $\varepsilon \rightarrow \gamma$ structural transition. Among the elements studied, Mn, Mo, V, and W exert the most pronounced elevating effects in ISFE, thereby extending the thermodynamic stability of the fcc phase at elevated temperatures. By contrast, Ni and Fe exhibit a comparatively marginal influence, while Cr and C act to elevate the transformation temperature.

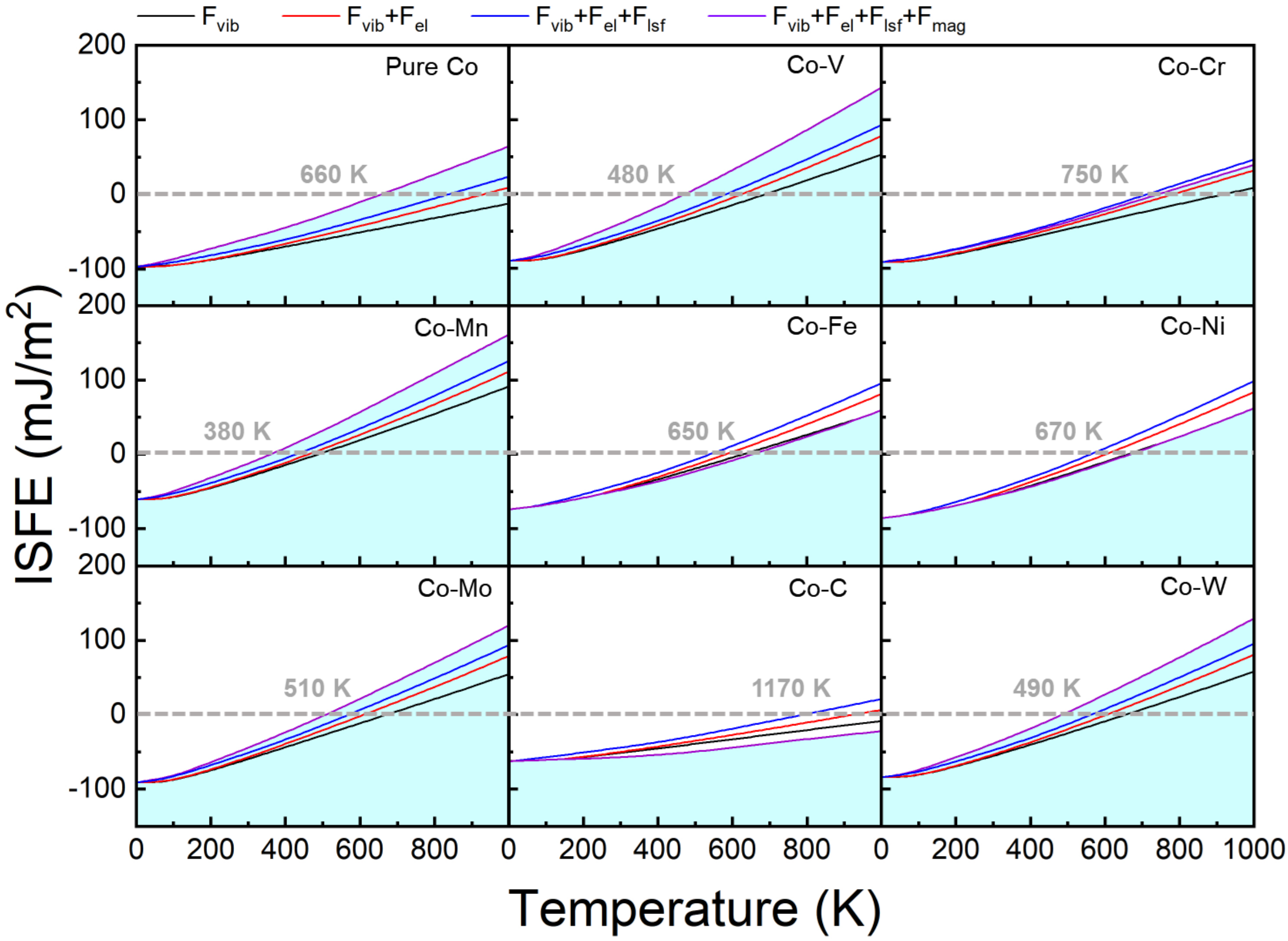

**Fig. 7.** Temperature dependence of the ISFE in Co-based alloys. The light blue region represents the total ISFE after accounting for all contributions. The intersection of the gray dashed line and the purple curve denotes the $\varepsilon \rightarrow \gamma$ transformation temperature.

To validate these predictions against established phase diagrams and experimental benchmarks, we categorize the binary cobalt systems based on their underlying thermodynamic complexity. The first cohort includes systems with substantial terminal solubility in proximity to the phase transition, therefore the phase stability is primarily governed by the direct competition between the $\gamma$ and $\varepsilon$ phases in their solid solution state. This group, exemplified by Co-Ni, Co-Mn, and Co-Cr, provides relatively direct benchmarks for our model. For instance, while both Ni and Mn act as $\gamma$ stabilizers that lower the transformation temperature, Ni demonstrates a considerably weaker stabilizing effect than Mn [51, 52], exerting almost no influence on the transition temperature at dilute concentrations. This observation aligns closely with our predictions, which show that 0.93 at.% Ni has only a negligible effect on the transformation temperature, whereas 0.93 at.% Mn reduces it to approximately 380 K. In contrast, Cr serves as a potent $\varepsilon$ stabilizer and raises the transition temperature in Co to a range between 720 and 765 K for a 0.83 at.% addition [51]. This is in excellent agreement with the current prediction of an elevated transformation temperature of approximately 750 K for a 0.93 at.% addition. Consequently, for systems governed primarily by phase competition in their solid solution state, the present model accurately captures both the sign and the relative magnitude of dilute solute effects on phase stability.

The second group encompasses thermodynamically intricate regimes, such as Co-V, Co-Mo, Co-W, Co-C, and, to some extent, Co-Fe. In these systems, the phase landscape near the transition temperature is significantly modulated by the emergence of discrete third phases, e.g., $Co_3X$ (X=W, V, Mo) intermetallics and precipitation of graphite or bcc Fe [51]. Consequently, the equilibrium terminal solubilities of these elements in Co are often negligible at the $\varepsilon \rightarrow \gamma$ transition temperature. Nevertheless, in realistic Co alloys, kinetically trapped supersaturated states are prevalent. The substantial solid solubility achieved at high temperatures is effectively preserved during cooling due to restricted diffusion kinetics, rendering equilibrium phase diagrams insufficient for capturing the solute effects of these elements.

A special case might be the Co-Fe system, where maximum terminal solubility of Fe in $\varepsilon$ phase Co is approximately 3 at.%. Within this dilute solid solution range, Fe is known to reduce the phase transition temperature to approximately 653 K at 1 at.% of solute [51], closely matching the value predicted in Fig. 7. Beyond this solubility limit, investigations of supersaturated Fe also provide crucial benchmarks. Tisone et al. [15] reported that Fe additions of 6-12 at.% increase the ISFE of Co. Specifically, an experimental ISFE of 11.9 mJ/m$^2$ was reported for Co-6 at.% Fe at 300 K. Linear extrapolation of our calculated value from 0.93 at.% to 6 at.% Fe gives 11.59 mJ/m$^2$, in excellent agreement with experiment. Although the low-temperature phase stability of Co-Fe may be affected by bcc-related equilibria, this direct energetic comparison bypasses the complexities of phase boundary assessment and thus provides a more robust quantitative benchmark for our model.

Overall, the present model not only captures the dilute-solute stabilization trends documented in experimental phase diagrams, but also facilitates a direct quantification of the intrinsic solute impact on ISFE within regimes typically obscured by third-phase formation, eutectoid reactions, or kinetic hysteresis. In contrast to empirical assessments derived from scattered transformation-temperature data, which are frequently confounded by kinetic effects, the present approach establishes a rigorous thermodynamic foundation for elucidating dilute alloying effects in cobalt alloys. By isolating the intrinsic thermodynamic drivers from extrinsic kinetic artifacts, this approach provides a predictive roadmap for the rational design of advanced materials with precisely controlled stacking fault energies and structural phase stability.

**4. Discussion**

Taken together, the present results clearly demonstrate that a comprehensive thermodynamic treatment of stacking fault energy is essential for capturing phase behavior. As evidenced in Figs. 5b and 7, V, Cr, W, and Mo all reduce the ISFE of Co at 0 K. From a purely static energetic perspective, these solute additions would be expected to stabilize the hcp phase and elevate the $\varepsilon \rightarrow \gamma$ transformation temperature. However, at finite temperatures, a striking reversal is observed. Only Cr raises the transformation temperature, whereas V, W, and Mo consistently suppress it and promoting fcc stability. This stark discrepancy underscores that alloying effects derived at 0 K defy direct extrapolation to high temperature phase stability. Such a failure arises because vibrational, electronic, and magnetic free energy contributions fundamentally reconfigure the stacking fault energetics and the structural equilibrium between the $\gamma/\varepsilon$ phases.

This point is especially profound for WC-Co cemented carbides. While W and C represent the primary chemical elements, their respective thermodynamic signatures in Co binder remain fundamentally asymmetric. Previous studies have shown that the dissolution and redistribution of W in the Co binder are highly dynamic during sintering: above 950 °C, the binder composition generally adheres to the equilibrium phase boundaries; however, during the cooling stage, the W concentration profile becomes kinetically frozen at around 800-900 °C [53]. In contrast, the carbon content retained in the binder is much narrower and strongly dependent on the cooling rate. Reported W contents in the Co binder span roughly 1-26 wt.%, whereas carbon is typically only about 0.1-0.4 wt.% and may become virtually undetectable following slow cooling [54].

To further clarify the effect of W content on the Co binder phase, we compared the microstructure and local composition of the two samples described in Section 2.4, namely the as-sintered specimen and the quench-tempered specimen. Typical TEM images of the Co phase in these two samples are shown in Fig. 8a1 and b1. We observe that the stacking-fault density in the as-sintered sample is markedly higher than that in the quench-tempered sample. At higher magnification, the stacking faults in the as-sintered sample are clearly wider than those in the quench-tempered sample and can even develop into extensive stacking-fault bands, accompanied by a more diffuse electron diffraction pattern (Fig. 8b2). In contrast, the quench-tempered sample exhibits much narrower stacking faults, including faults that terminate at both ends within the grain (Fig. 8a3). Meanwhile, EDS results reveal that the W content in the Co phase (C was excluded from the calculation) is significantly lower in the as-sintered state (1.89 at.%) than in the quench-tempered state (2.48 at.%). This difference arises because the rapid cooling rate during quenching effectively retains the highly concentrated W solid solution formed at elevated temperatures. Based on these microstructural correlations, we infer that an increase in W content suppresses the formation of stacking faults by increasing the stacking-fault energy. This interpretation is consistent with both the experimental results reported by Zhou et al. [17] and our thermodynamic calculations shown in Fig. 7.

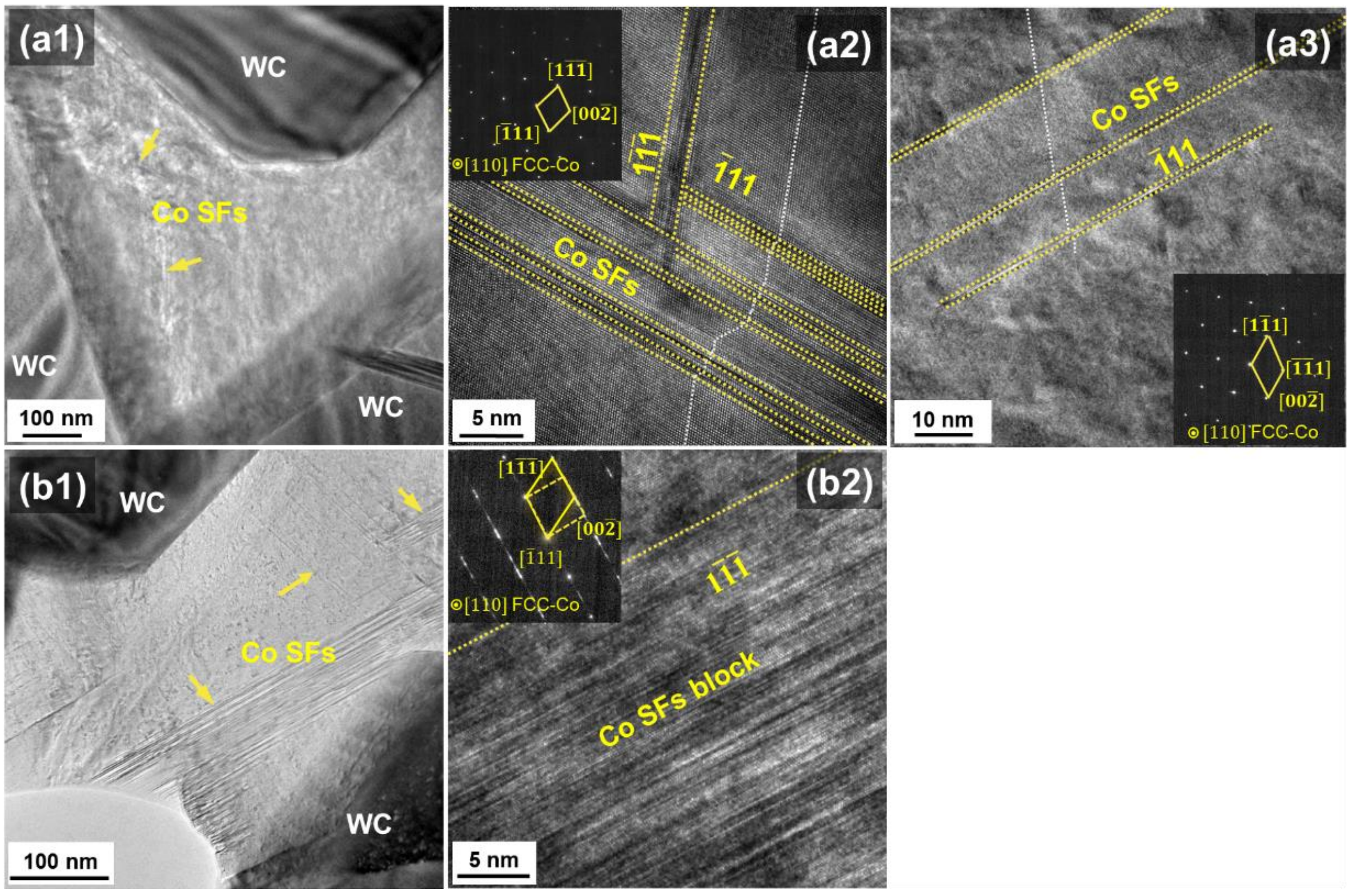

**Fig. 8.** Microstructures of WC-Co cemented carbide in the quench-tempered (a) and as-sintered (b) specimen. Yellow

dashed lines indicate stacking-fault regions.

Beyond phase stability, the evaluated thermal evolution of ISFE suggest distinct plastic deformation mechanisms among the alloyed systems. In fcc crystals, SFE directly controls the dissociation width of partial dislocations, the probability of cross-slip, and the tendency for deformation twinning. In general, a decrease in SFE shifts the dominant deformation mode from conventional dislocation glide to twinning, and eventually to deformation modes assisted by stacking faults or the $\gamma \rightarrow \varepsilon$ transformation. Experimental work on pure Co has also shown that retained fcc phase at room temperature alters the mechanical response, and that the allotropic transformation in polycrystalline Co can be coupled with plastic deformation, with basal slip competing with twinning during the transformation [55]. Consequently, we infer that Cr, which stabilizes $\varepsilon$ phase and increases the transformation temperature, may promote stacking-fault accumulation within the fcc Co matrix, thereby enhancing the work-hardening capability during deformation. In contrast, solutes like W, Mo, and Mn increase the ISFE at elevated temperatures. These additions likely suppress the formation of stacking faults and hcp embryos, thereby shifting plastic deformation toward more conventional dislocation-mediated slip. Nevertheless, this interpretation remains a physically motivated extrapolation, as definitively resolving the competition among slip, twinning, and phase transformation requires evaluating the full generalized stacking fault energy landscape at finite temperatures, which represents a critical avenue for future research.

## 4. Conclusions

This work integrates first-principles thermodynamic calculations with experimental characterization to elucidate the effects of alloying elements on the intrinsic stacking fault energy (ISFE) and phase stability of Co-based alloys. The main conclusions are summarized as follows:

(1) At 0 K, the influence of transition-metal solutes on the ISFE of Co is primarily dictated by atomic-size effects. Excluding specific 3d elements with dominant magnetic signatures, a robust linear correlation exists between the solute induced volume strain and the change in stacking fault energy. This identifies atomic misfit volume as the governing factor for stacking fault energetics in the dilute limit.

(2) We show that static energy calculations at 0 K are insufficient to accurately describe alloying effects on phase stability, as they frequently fail to capture the correct sign or magnitude of solute contributions. By incorporating vibrational, electronic, longitudinal spin-fluctuation, and magnetic free-energy contributions, the present thermodynamic framework reconciles these discrepancies and accurately captures the temperature dependent stacking fault energetics of Co and its alloys. The predicted stabilization trends across diverse alloying systems align closely with established binary phase diagrams, highlighting that a rigorous thermodynamic treatment is indispensable for aligning atomistic predictions with experimental observations.

(3) TEM/EDS characterization of WC-Co cemented carbides reveals that a higher W content in the Co binder phase significantly suppresses stacking-fault formation. This observation is consistent with the finite-temperature thermodynamic prediction that W increases the stacking fault energy of Co, providing a direct link between atomic scale thermodynamics and the observed microstructural evolution.

Overall, this work establishes a physically grounded framework for analyzing stacking fault energy and phase stability in Co-based alloys, providing theoretical guidance for tailoring phase stability and deformation behavior in Co-based alloys and WC-Co cemented carbides.

**Acknowledgements**

This work was financially supported by National Natural Science Foundation of China (No.: 52401010; 12375260), Hunan Provincial Natural Science Foundation of China (No.: 2025JJ40005). State Key Laboratory of Cemented Carbide Construction Project (No.: 2024ZYT006). State Key Laboratory Project of China Minmetals Corporation (No.: 2025GZYJ01). We acknowledge Hefei Advanced Computing Center for providing computing resources.

**Competing interests**

The authors declare no competing interests.

**Data availability**

The data generated and/or analyzed within the current study will be made available upon reasonable request to the authors.

**Author contributions**

**Zheng Zhong:** Methodology, Formal analysis, Investigation, Data Curation, Writing - Original Draft, Visualization. **Ziqi Cui:** Methodology, Formal analysis, Investigation, Data Curation. **Yu Zhuo:** Writing - review & editing. **Tianyu Yu:** Writing - review & editing. **Jianfeng Cai:** Writing - Review & Editing. **Kaibo Zou:** Resources, Writing - review & editing. **Jiacheng Shen:** Writing - review & editing. **Bowen Huang:** Writing - review & editing. **Zhuoming Xie:** Writing - review & editing. **Huiqiu Deng:** Writing - review & editing. **Yang Yu:** Writing - review & editing. **Hao Zhang:** Resources, Supervision, Writing - review & editing. **Wangyu Hu:** Resources, Supervision, Writing - review & editing. **Tengfei Yang:** Writing - review & editing. **Jie Hou:** Conceptualization, Methodology, Validation, Formal analysis, Resources, Data curation, Writing - Review & Editing, Supervision, Project administration, Funding acquisition.